\documentclass{ws-ijmpa}

\begin{document}

\newcommand{\drbar}{{\overline{DR}}}
\newcommand{\msbar}{{\overline{MS}}}
\newcommand{\GeV}{{\rm GeV}}

\newcommand{\stau}{\tilde{\tau}}
\newcommand{\snt}{{\tilde{\nu}_\tau}}
\newcommand{\ur}{\tilde{u}_R}
\newcommand{\ul}{\tilde{u}_L}
\newcommand{\dr}{\tilde{d}_R}
\newcommand{\dl}{\tilde{d}_L}
\newcommand{\st}{\tilde{t}}
\newcommand{\sbot}{\tilde{b}}
\newcommand{\sg}{\tilde{g}}
\newcommand{\nt}{\tilde{\chi}^0}
\newcommand{\cp}{\tilde{\chi}^+}
\newcommand{\cm}{\tilde{\chi}^-}
\newcommand{\cx}{\tilde{\chi}}
\newcommand{\ser}{\tilde{e}_R}
\newcommand{\sel}{\tilde{e}_L}
\newcommand{\smul}{\tilde{\mu}_L}
\newcommand{\smur}{\tilde{\mu}_R}

\newcommand{\sne}{\tilde{\nu}_e}
\def\lsim{\mathrel{\raise.3ex\hbox{$<$\kern-1.05em\lower1ex\hbox{$\sim$}}}}
\def\gsim{\mathrel{\raise.3ex\hbox{$>$\kern-1.05em\lower1ex\hbox{$\sim$}}}}
\def\ie{{\it i.e.}}
\def\eg{{\it e.g.}}
\def\etc{{\it etc.}}
\def\vs{{\it vs.}}
\def\opcit{{\it op.~cit.}}

\markboth{Jan Kalinowski}
{Supersymmetry at and beyond the LHC}

%
\catchline{}{}{}{}{}
%

\title{SUPERSYMMETRY AT AND BEYOND THE LHC}

\author{JAN KALINOWSKI}

\address{Institute of Theoretical Physics, University of Warsaw\\
Hoza 69, 00681 Warsaw, Poland\\
jan.kalinowski@fuw.edu.pl}

\maketitle

\begin{history}
\received{Day Month Year}
\revised{Day Month Year}
\end{history}

\begin{abstract}
Prospects for SUSY discoveries and measurements at
future colliders LHC and ILC are discussed. The problem of reconstructing the underlying
theory and SUSY breaking mechanism is also addressed.

\end{abstract}

\ccode{PACS numbers: 12.15.-y, 12.60.-i}

\section{Introduction}
The Standard Model (SM) is very successful in describing the constituents of matter and their interactions at and below the electroweak scale\cite{ewwg}. However, it does not address many  important issues, like the mass generation and mass pattern,  the unification of all forces (including gravity), the matter composition of our universe etc. These issues seem to point to  new phenomena  at a TeV scale which can experimentally be tested soon at the Large Hadron Collider (LHC) and in (hopefully) not too far a future at the International Linear Collider (ILC).

Although the answers to these issues could have different origin, it is very tempting to contemplate supersymmetry\cite{susy} (SUSY) as responsible for all of them. SUSY turned to be able to beautifully accommodate or explain (at least in the technical sense) some of the SM problems, {\it e.g.} it solves the hierarchy problem, explains the gauge coupling unification, provides the radiative electroweak symmetry breaking, provides a candidate for dark matter (DM), offers new ideas on matter-antimatter asymmetry of the universe \etc\ SUSY still lacks any direct experimental evidence, however, is not yet excluded either.

Discovering supersymmetry,
the main candidate for a unified theory beyond the SM,
is the challenge for world physics community experimenting at existing and
future colliders.  Many detailed phenomenological studies of SUSY at present
and future colliders have been performed in the past. Here
only some selected results are presented  on the discovery potentials of the main two LHC detectors: ATLAS and CMS. Assuming that SUSY is discovered at LHC we will
discuss how experimentation at the ILC will help in revealing the
details of the underlying model and  address the question of
reconstructing the fundamental SUSY parameters and the mechanism of SUSY breaking.

\section{Supersymmetry searches at the LHC}
At present the most restrictive limits on the SUSY parameter space
come from negative results of SUSY searches\cite{tevhera} at two colliders: Tevatron at
Fermilab and HERA at DESY. Both machines perform beautifully
and significant
improvements (or discoveries) can be expected in near future until the LHC will start taking data.

The strongly interacting squarks and gluinos ($\tilde{q}$ and $\tilde{g}$), if they are in the TeV range, will be copiously produced at the LHC with production cross sections comparable to jet production with transverse momenta $p_t\sim$ SUSY masses (typically in the picobarn range).  Direct production of weakly interacting sparticles has much lower rates.  Squarks and gluinos will promptly decay into jets and lighter SUSY particles which will further decay. Generically one can expect in the final state high-$p_t$ jets and leptons, possibly large missing energy $\not \!\!E_t$, or displaced vertices \etc\  Since the LHC detectors are designed to detect these objects, they are well equipped to cover a broad spectrum of possible decay modes of SUSY particles. There have been many experimental analyses demonstrating the capabilities of the LHC detectors ATLAS and  CMS and we refer to technical design reports\cite{ATLAStdr,CMStdr} of both collaboration for more details.

\subsection{Inclusive searches  at LHC}
\begin{figure}[h!]
\centering
\includegraphics[height=4cm,width=4cm]{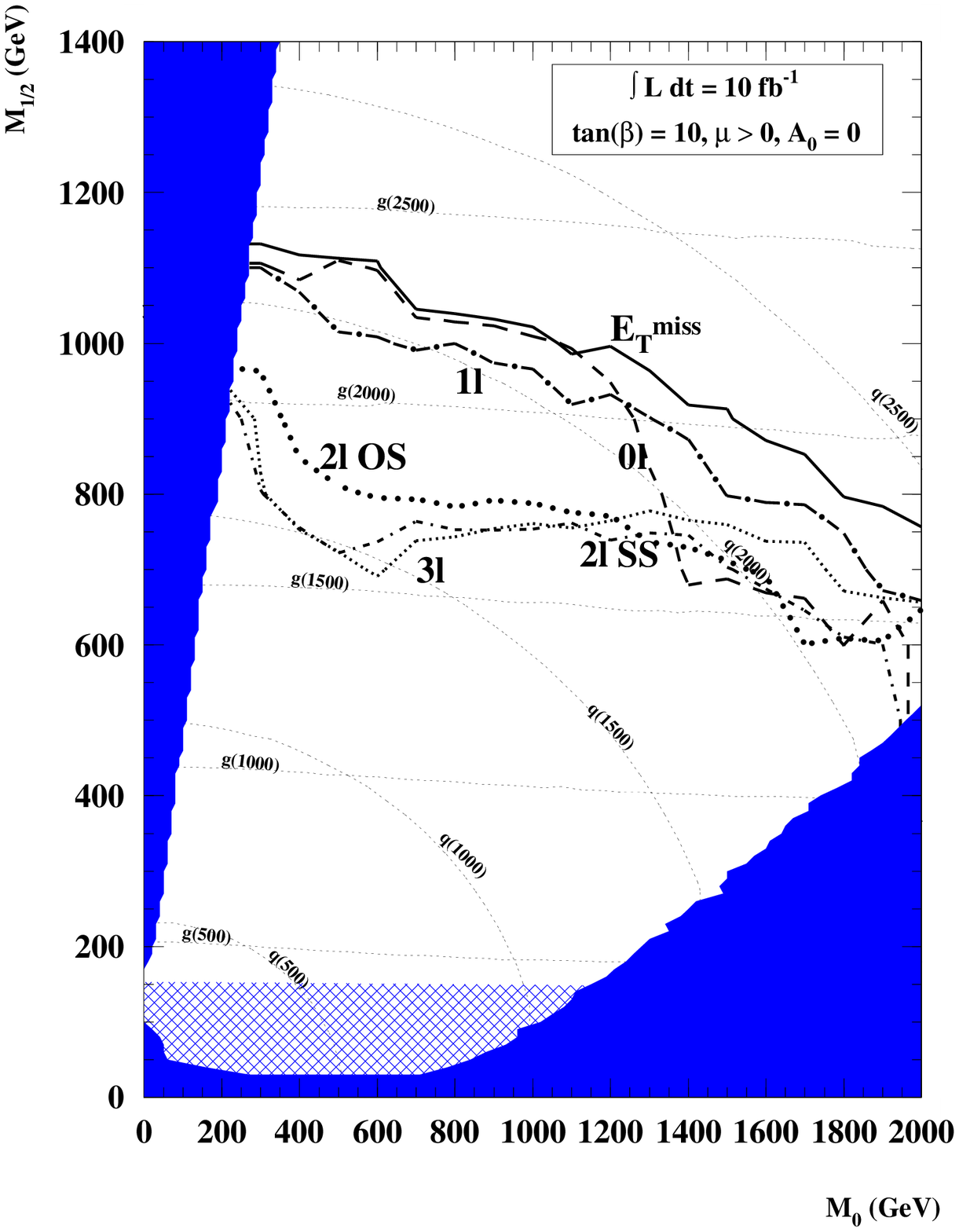}
\includegraphics[height=4cm,width=4cm]{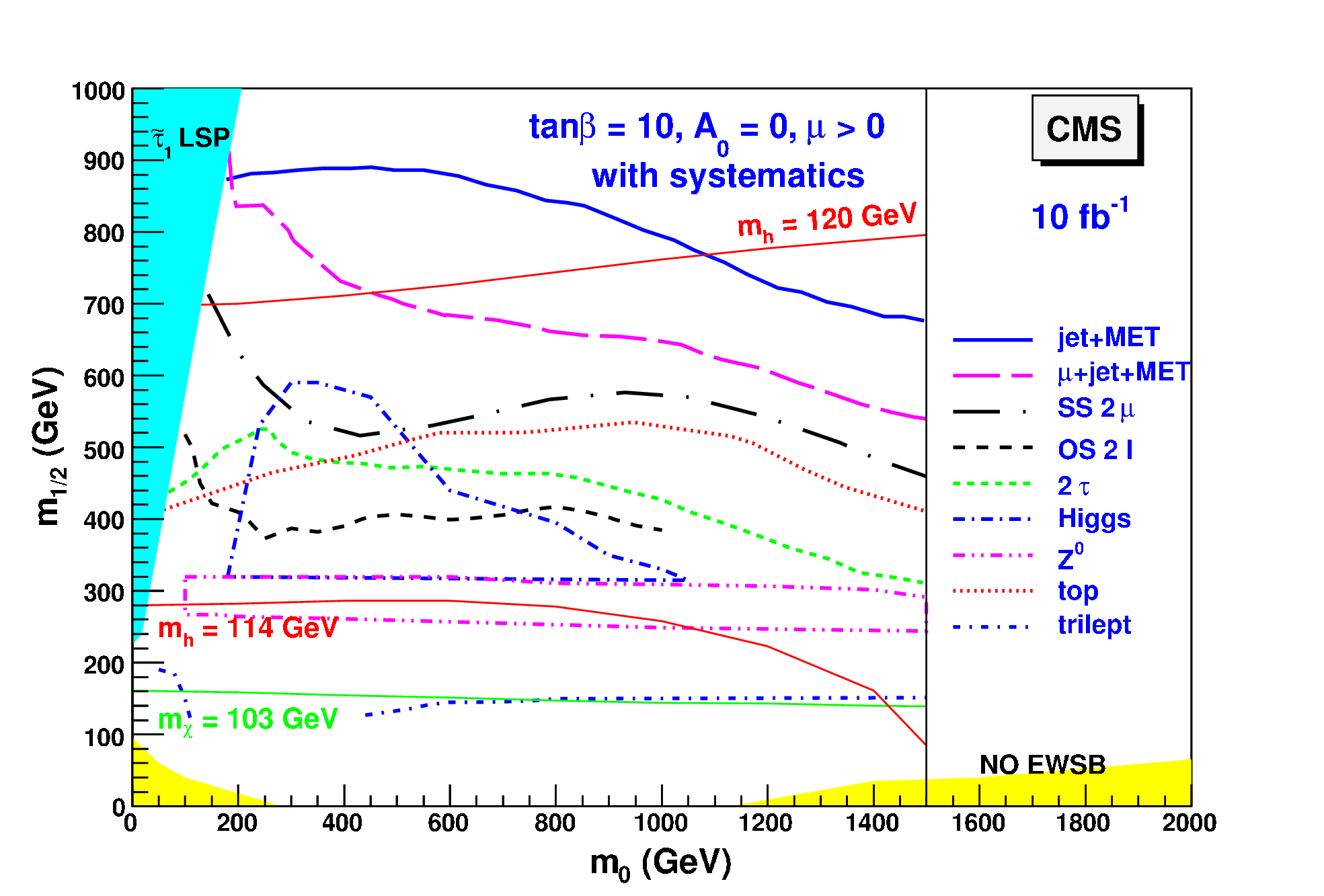}
\includegraphics[height=4cm,width=4cm]{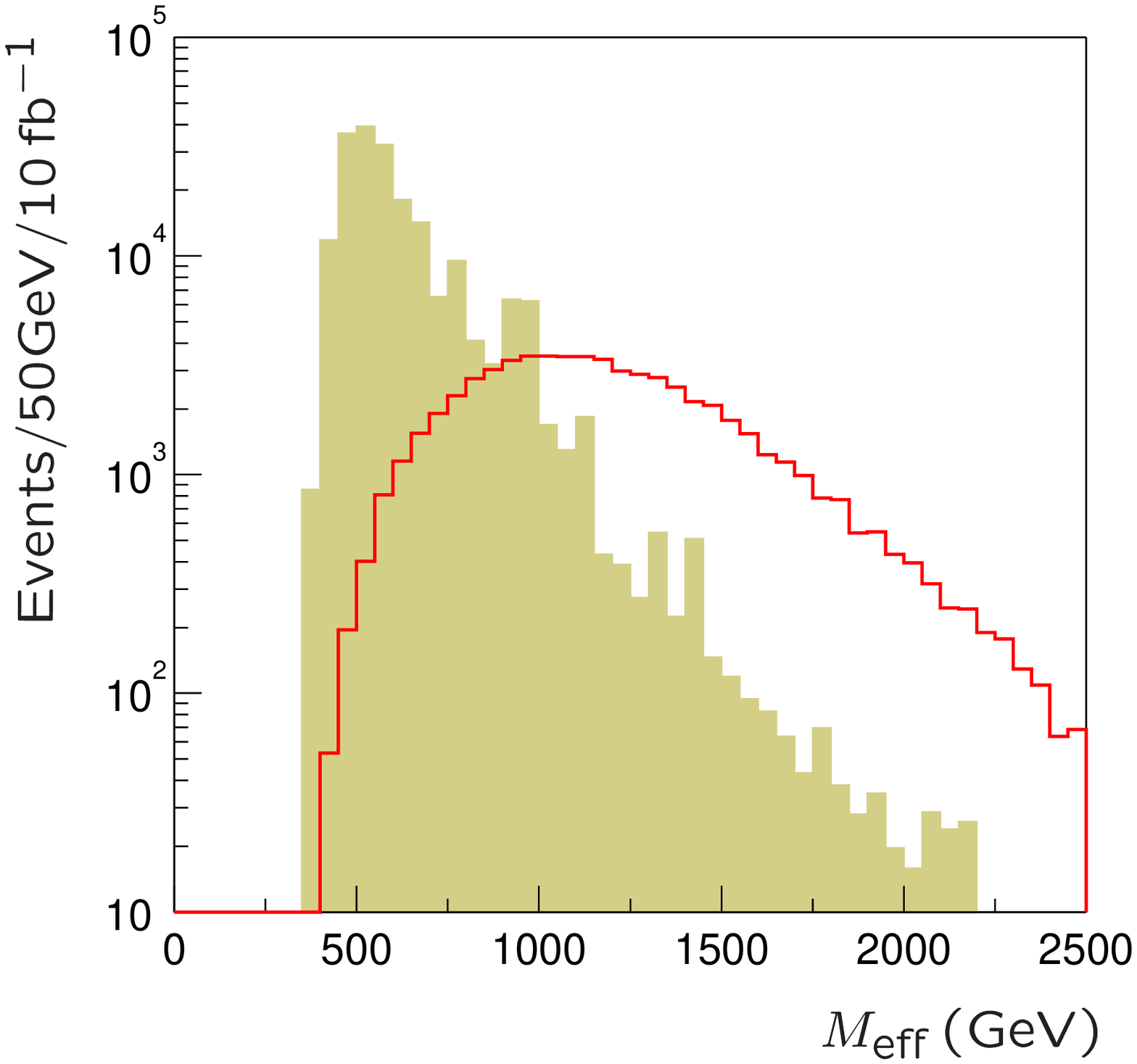}
  \caption{\it
Search limits for various channels in the mSUGRA
    parameter space (left and center) and  $M_{\rm eff}$ distribution for a
    mSUGRA point and SM background after cuts (right).}
\label{fig:inclusive}
\end{figure}

Jets from squark and gluino decays will have large transverse momenta $p_t$ of the order of sparticle masses. If the lightest SUSY particle (LSP) is stable, as in scenarios with $R$-parity conserved, it will escape undetected giving large $\not \!\!E_t$. The SM background events from top quark, $W$ and $Z$ boson decays  do not have such high-$p_t$ objects. A set of simple cuts can then be designed to enhance the signal over the background in inclusive ``transverse'' searches for SUSY particles.  For example, in typical mSUGRA scenarios, requiring at least four jets with large $p_t^i$ and large
\begin{eqnarray}
M_{\rm eff}=\sum_{i=1,\ldots 4} p^i_t +\not \!\!E_t
\end{eqnarray}
and selecting events spherical in the transverse plane
(specific cuts  depend on details of
the model) can be sufficient to discover new particles\cite{ATLAStdr}. To reduce the
background further, hard, isolated lepton(s) may be required and their $p_t$
is then included in the definition of $M_{\rm eff}$.
The reach of inclusive searches at 10$^{-1}$ fb is illustrated in
Fig.~\ref{fig:inclusive}; and squarks and gluinos with masses up to $\sim 2.5$
TeV can be found at LHC with 100 fb$^{-1}$.
Monte Carlo studies have also shown that the position of the peak
in $M_{\rm
  eff}$ distribution  correlates quite well with sparticle masses, namely
$M_{\rm eff} \sim {\rm min}(m_{\tilde{q}},m_{\tilde{g}})$,
providing a first estimate of the overall SUSY mass scale,
Fig.~\ref{fig:inclusive} right panel.

Recently the importance of including exact matrix
element corrections to the previous parton shower estimate of the background has been emphasized\cite{exact}. This
may significantly change the background distribution in the signal region. This is particularly  important in scenarios with
sparticle masses degenerate in which the signal events are less ``transverse''.
As a result, the standard SUSY cuts reduce the signal sample and SUSY discovery
is more affected by the SM background. Such a scenario occurs, for example,
in a string inspired model based on the flux compactification\cite{KKLT}, in which the unification scale of the
soft SUSY parameters  can be much lower than the GUT scale, even
of the order of of the weak scale\cite{mmam}.  Depending on the ratio of F-terms
of the volume modulus field and the mSUGRA
compensator field, the
mass spectrum of SUSY particles changes smoothly from the mSUGRA--like to
the anomaly-mediation--like. There are regions of parameters where the squark, slepton and gaugino masses are significantly degenerated.
If $m_{\tilde{\chi}^0_1}
\gsim m_{\tilde{q},\tilde{g}}/2$,   the signal $M_{\rm eff}$ distribution
becomes quite    similar to that of the background. New ideas are needed to improve search strategies. For example, examining the pattern of events in the $M_{\rm eff}$-$\not \!\!E_t$ plane may help to discriminate signal from background better\cite{KN}.

\subsection{Sparticle mass measurements}
In $R$-parity conserving SUSY all sparticles decay into invisible LSP, so no
mass peaks can be directly reconstructed. Nevertheless, it might be possible to identify
particular decay chains and exploit the ``endpoint method'' to measure
combinations of masses\cite{endpoint}. A relatively clean
channel, for example,  is provided by the three-body decay or, if the slepton can be
on-shell,
the cascade of two-body decays of
the heavier neutralino
\begin{eqnarray}
\tilde{\chi}^0_i\to \tilde{\ell}\ell\to \ell\ell\tilde{\chi}^0_1
\end{eqnarray}
The di-lepton mass distribution endpoints
are functions of the masses of sparticles involved in the decay
\begin{eqnarray}
m_{\ell\ell}({\mbox{3-body}})&=& m_{\tilde{\chi}^0_i}-m_{\tilde{\chi}^0_1}\\
m_{\ell\ell}({\mbox{2-body}})&=& \sqrt{(m^2_{\tilde{\chi}^0_i}
  -m^2_{\tilde{\ell}})
 (m^2_{\tilde{\ell}}- m_{\tilde{\chi}^0_1})}/m_{\tilde{\ell}}
\end{eqnarray}
Requiring two isolated leptons in addition to
multi-jet and $\not\!\!E_t$ cuts, like those described above, the signal events can be selected. If lepton flavor
is conserved, contributions from two uncorrelated decays
cancel in the combination of $e^+e^-+\mu^+\mu^--e^\pm\mu^\mp$ sample giving a very
clean signal and allowing  a precise endpoint measurement. The shape of the
distribution also helps to distinguish two-body from  three-body decays.
\begin{figure}[h!]
\includegraphics[height=.2\textheight,width=.3\textwidth]{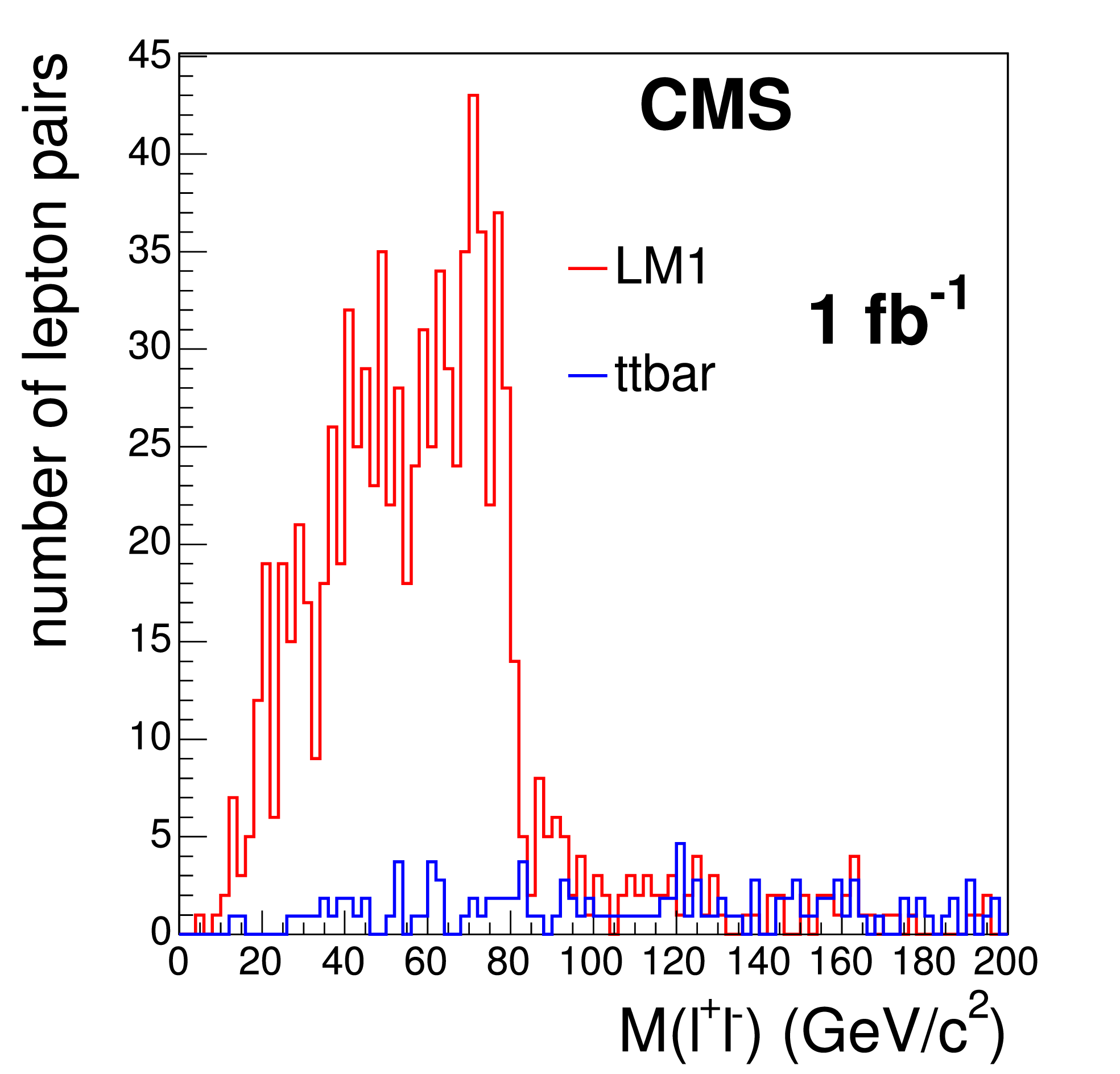}
\includegraphics[height=.2\textheight,width=.33\textwidth]{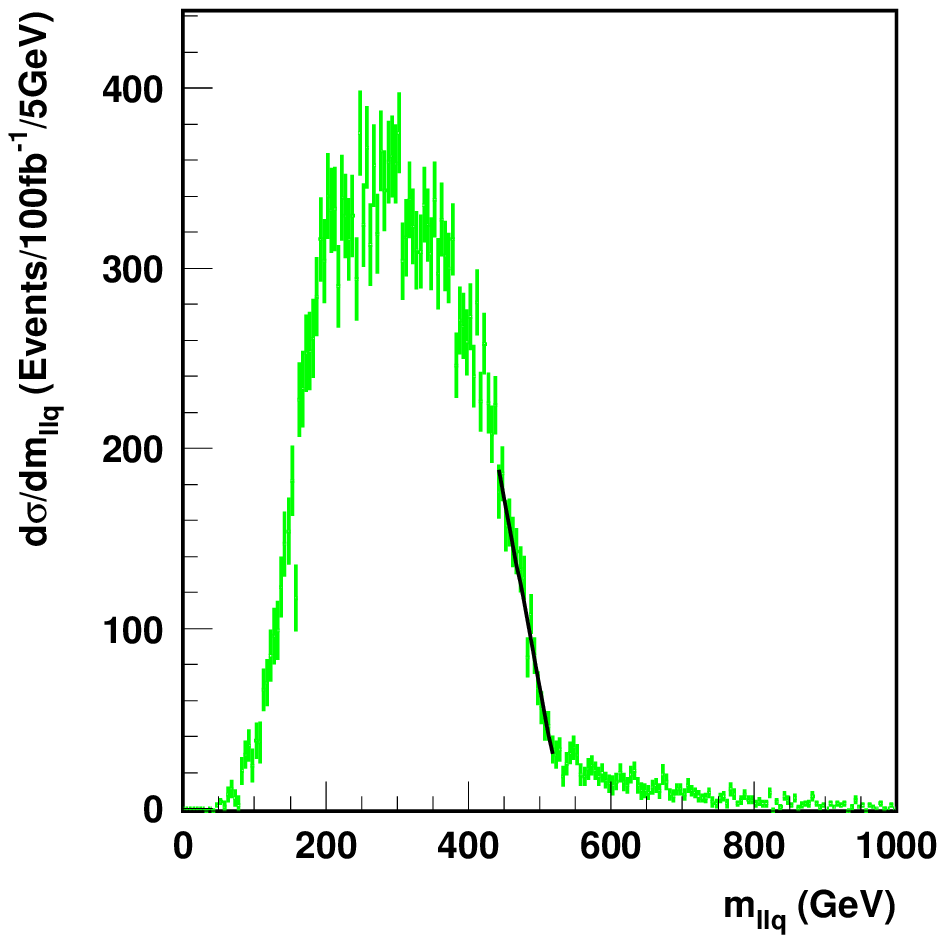}
~~\begin{minipage}[b]{4cm}
  \caption{\it Di-lepton (left) and $\ell\ell q$ (right)
invariant mass distributions from
    $\tilde{\chi}^0_2$ cascade decay.
    \label{fig:edges} }\end{minipage}
\end{figure}

\noindent
Long decay chains, like
\begin{eqnarray}
\tilde{g}\to\ j_1\tilde{q}\to\tilde{\chi}^0_2 j_1 j_2\to
\tilde{\ell}\ell_1 j_1 j_2\to \tilde{\chi}^0_1 \ell_1\ell_2 j_1 j_2
\label{eq:longchain}
\end{eqnarray}
expected in some mSUGRA scenarios \eg\  SPS1a'\cite{SPA}, allow more endpoint measurements. With two jets and two leptons in the final state it should be possible to measure the endpoints of invariant mass distributions $\ell\ell$, $\ell\ell j$, $\ell j$, like those shown in Fig.~\ref{fig:edges}.   Although these endpoints are smeared by jet reconstruction, hadronic resolution, and miss-assignment of the jets that come from squark decays, these endpoints should be measured at the level of 1\%, \ie\ determining mass relations to 1-2\%\cite{gjelsten}. In fact, with so many endpoints one can solve for the absolute values of the unknown masses of ${\tilde{g}}$, ${\tilde{q}}$, ${\tilde{\chi}^0_2}$, $\tilde{\ell}$ and ${\tilde{\chi}^0_1}$ within 5--10\% accuracy. This is a general feature of the determination of sparticle masses when the LSP momentum cannot be measured directly. For this particular point, already  ${\cal O}(5)$\% accuracy in the mass of sleptons and the lightest neutralino can provide a link to cosmology. Based in this information one can calculate the neutralino annihilation rate at the time of decoupling and estimate the amount of DM at the level of 7\% \cite{DMcalc}. For other scenarios, however, the expected accuracy can be much worse\cite{Baltz:2006fm}.

It is notable that via the above decay chain the LHC can access the heaviest neutralino
$\tilde{\chi}^0_4$ which in the SPS1a' scenario is too heavy to be
produced at the 500 GeV $e^+e^-$ collider. The measured mass difference
$m_{\tilde{\chi}^0_4}-m_{\tilde{\chi}^0_1}$,
in the same decay chain as in eq.(\ref{eq:longchain}), but with
$\tilde{\chi}^0_4$ replacing $\tilde{\chi}^0_2$, would provide an important constraint
on model parameters.   If the measurements at the LHC and ILC could be
combined the errors for the MSSM Lagrangian parameters would
significantly  be reduced\cite{Desch:2003vw}.

The mass determination through the endpoint method has several shortcomings:
the LSP momentum cannot be reconstructed except for a few very special points
in the parameter space, only events near endpoints are used neglecting
independent information contained in  events away, and the selected events may
contain contributions from several cascade decays causing additional systematic
uncertainties. An alternative  ``mass
relation'' method\cite{mass-rel}, which exploits the on-shell conditions for
sparticle masses in the decay chains, allows  to solve for the kinematics and
reconstruct the SUSY masses as peaks in certain distributions. For example,
in the cascade decay eq.(\ref{eq:longchain}) five on-shell conditions can be
written for $\tilde{g}$,
${\tilde{q}}$, ${\tilde{\chi}^0_2}$, $\tilde{\ell}$
and ${\tilde{\chi}^0_1}$ in terms of the measured
momenta of leptons, jets and 4 unknown momentum
components of the undetected neutralino. Each event, therefore, spans
a 4-dim hypersurface in a 5-dim mass space, and in principle 5 events would be
enough to solve for masses of involved sparticles. Note that events need not
be close to endpoints of the decay distributions, \ie\ the method can be used
even if the number of signal events is small.

\subsection{Proving it is SUSY}
A generic signal of large $\not \!\!E_t$, as in the weak-scale SUSY, arises in almost any model with the lightest ${\cal O}$(100 GeV) particle stable and neutral, as suggested by the dark matter of the universe. Therefore, we have to be able to distinguish the SUSY decay chain eq.(\ref{eq:longchain}) from, \eg, the cascade decay 
\begin{eqnarray}
{g}'\to\ j_1{q}'\to Z' j_1 j_2\to
{\ell}'\ell_1 j_1 j_2\to \gamma' \ell_1\ell_2 j_1 j_2
\label{eq:ued}
\end{eqnarray}
that arises in the universal extra-dimension model (UED)\cite{ued}. Here the
primes
denote the first excited Kaluza-Klein states of the corresponding
SM particles. In both cases the final state is the same
$\ell_1\ell_2 j_1 j_2$ with either the $\tilde{\chi}^0_1$ or the $\gamma'$
escaping detection.
What differentiates the
decays in eqs.(\ref{eq:longchain},\ref{eq:ued}) is the spins of intermediate
states  and the chiral structure of couplings. In contrast to the
UED case, in many processes the SUSY particles
are naturally   polarized due to the chiral structure of the theory.
For example, in the decay
 $\tilde{q}_L\to\tilde{\chi}^0_2 q_L$ the $\tilde{\chi}^0_2$ is polarized as
right-handed, opposite to $q_L$, because the $\tilde{q}\tilde{\chi}q$ Yukawa
coupling flips chirality.  The polarized neutralino further decays into either
$\tilde{\ell}_R\ell^+$ or $\tilde{\ell}^*_R\ell^-$ with equal rates (because
of the Majorana character of neutralinos), but  due to the chiral nature of
the Yukawa $\tilde{\ell}\tilde{\chi}\ell$ coupling, the $\ell^+$ is likely
to fly in the neutralino direction in the squark rest frame, while the $\ell^-$
in the direction of the quark jet. The difference in the angular distribution
is reflected as a charge asymmetry in the invariant mass distribution of  the
jet-lepton system\cite{Barr:2004ze}.

\begin{figure}[h!]
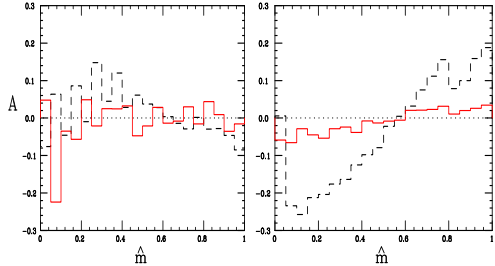

\centering
\includegraphics[height=.18\textheight,width=.25\textwidth]{ued_asym_det.ps}
\includegraphics[height=.18\textheight,width=.25\textwidth]
{sps_asym_det.ps}
~~~~\begin{minipage}[b]{50mm}
\caption{\it Detector-level charge asymmetries with respect to the jet+lepton
  rescaled invariant
mass, for UED- (left) and  SUSY-like (right) mass spectra. Dashed:
SUSY. Solid/red: UED. \label{fig:asymm}}\end{minipage}
\end{figure}
\noindent Although the charge asymmetry for $\tilde{q}^*_L$ decay is just opposite,  in
$pp$ collisions more squarks than anti-squarks are expected and the
$\tilde{\chi}^0_2$ production from squark decays is dominant. The amount of
charge
asymmetry in the $m(j\ell)$ is model dependent, Fig.~\ref{fig:asymm},
nevertheless it may  allow  resolving
the fermionic nature
of  the neutralino from the vector nature of the $Z'$ and confirm the
chiral structure of couplings\cite{smillie,plusbarr}.

\subsection{The LHC inverse problem}
The LHC experiments in the
supersymmetric particle sector offer not only the discovery potential but also
many high precision measurements of masses and
couplings.  The next step
towards establishing SUSY is the
reconstruction of low-energy SUSY breaking Lagrangian parameters
without assuming a specific scenario. This is a highly non-trivial task
\cite{Zerwas:2002as}.
In some favorable cases it might be possible to reconstruct the model. However, in many cases one is left with degenerate solutions, \ie\ many models could fit the LHC data equally well\cite{inverse}.

This task can be greatly ameliorated by experimenting
at the ILC where the experimental accuracies at the per-cent down to the
per-mil level are expected\cite{LCreports}.

\section{SUSY studies  at the ILC}

If the superpartner masses (at least some of them)
are in the TeV range, LHC will certainly see SUSY.
Many  different channels, in particular
from squark and gluino decays will be explored and many
interesting quantities measured, as discussed in the previous chapter.
However, to achieve
the ultimate goal of  all experimental efforts  to
unravel the SUSY breaking mechanism and
shed light on physics at high (GUT?, Planck?) scale, an   $e^+e^-$ LC
would be an indispensable tool\cite{LCreports}. First,
the LC will provide
independent checks of  the LHC findings. Second, thanks to the LC  unique
features: clean environment, tunable collision energy,
high luminosity, polarized
incoming beams, and possibly $e^-e^-$, $e\gamma$ and
$\gamma\gamma$ modes, it will offer
precise measurements of masses, couplings, quantum numbers,
mixing angles, CP phases etc. Last, but not least, it will
provide additional experimental input to the LHC analyses, like the
mass of the LSP.
Coherent analyses of data from the LHC {\it and} LC would thus allow for a
better, model independent reconstruction of low-energy  SUSY
parameters, and  connect  low-scale phenomenology with the high-scale
physics\cite{Weiglein:2004hn}.

An intense R\&D process and physics studies
since 1992 has lead to world-wide consensus that the
next high energy machine after the LHC should be an International Linear
Collider (ILC).
Planning, designing and funding the ILC requires global
participation and global organization. Therefore the Global
Design Effort for the ILC \cite{gde}, headed by Barry Barish,
has been established with the goal of preparing the project to be ready for
approval around 2010 and beginning construction around 2012. Recently released the
Reference Design Report\cite{RDR} defines the ILC baseline as follows:\\
\phantom{m} -  CM energy adjustable  from 200 to 500 GeV, and at $M_Z$ for
calibration,\\
\phantom{m} -  integrated luminosity of at least 500 fb$^{-1}$ in first 4
years,\\
\phantom{m} -  beam energy stability and precision below 1\%,\\
\phantom{m} - electron beam polarization of at least 80\%,\\
\phantom{m} - upgradeability to CM energy of 1 TeV.\\
The choice of options, like GigaZ (high luminosity run at $M_Z$), positron
polarization,
$e^-e^-$, $e\gamma$ or $\gamma\gamma$,  will depend on LHC+ILC physics
results.

Many detailed physics calculations and simulations have been performed and
presented during numerous
ECFA, ACFA and ALCPG workshops and LCWS conferences\cite{conferences}. Below
only some highlights are presented.

\subsection{Mass measurements}
\begin{figure}[htb]
\centering
\includegraphics*[width=35mm,height=35mm]{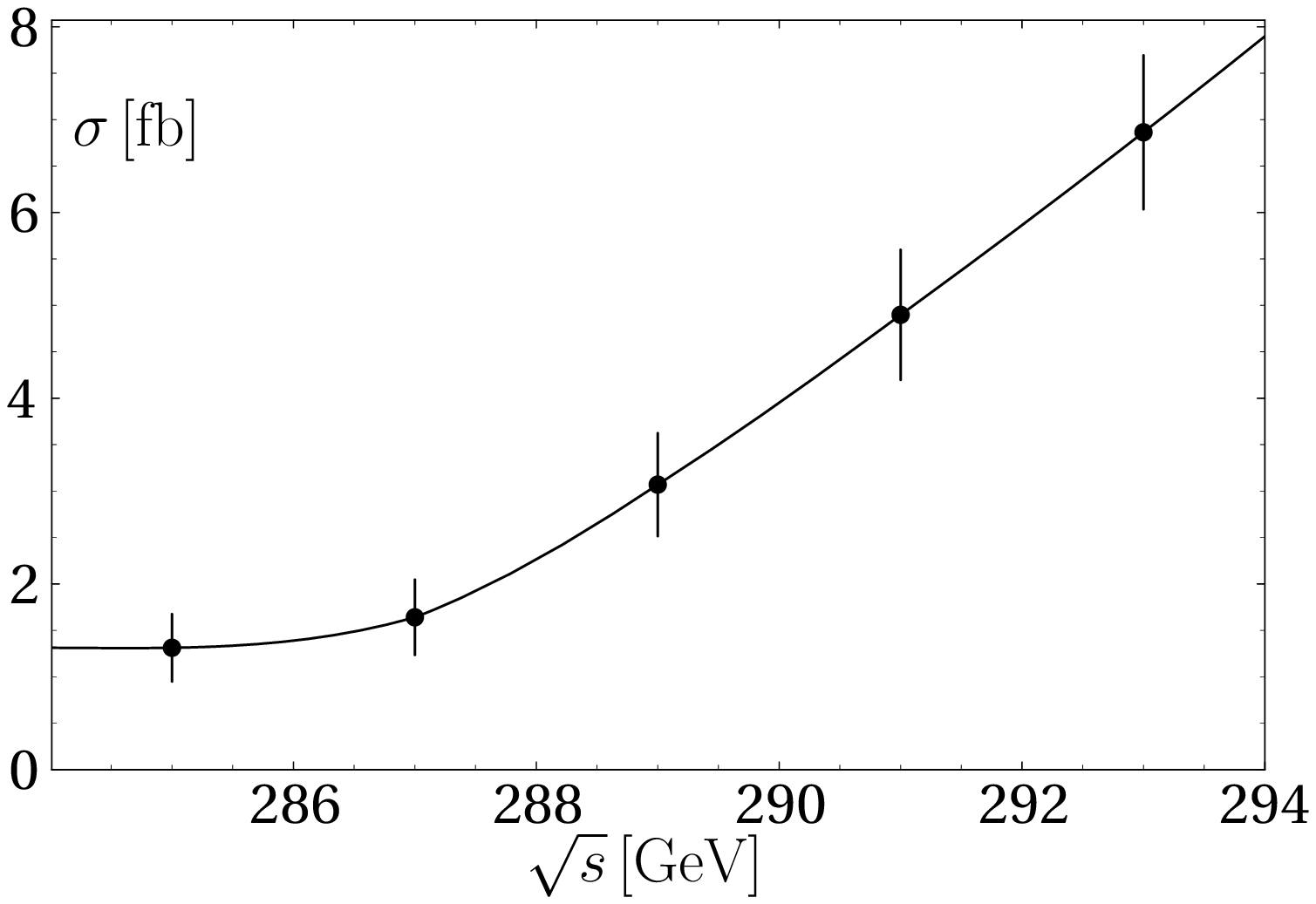}%
\includegraphics*[width=35mm,height=35mm]{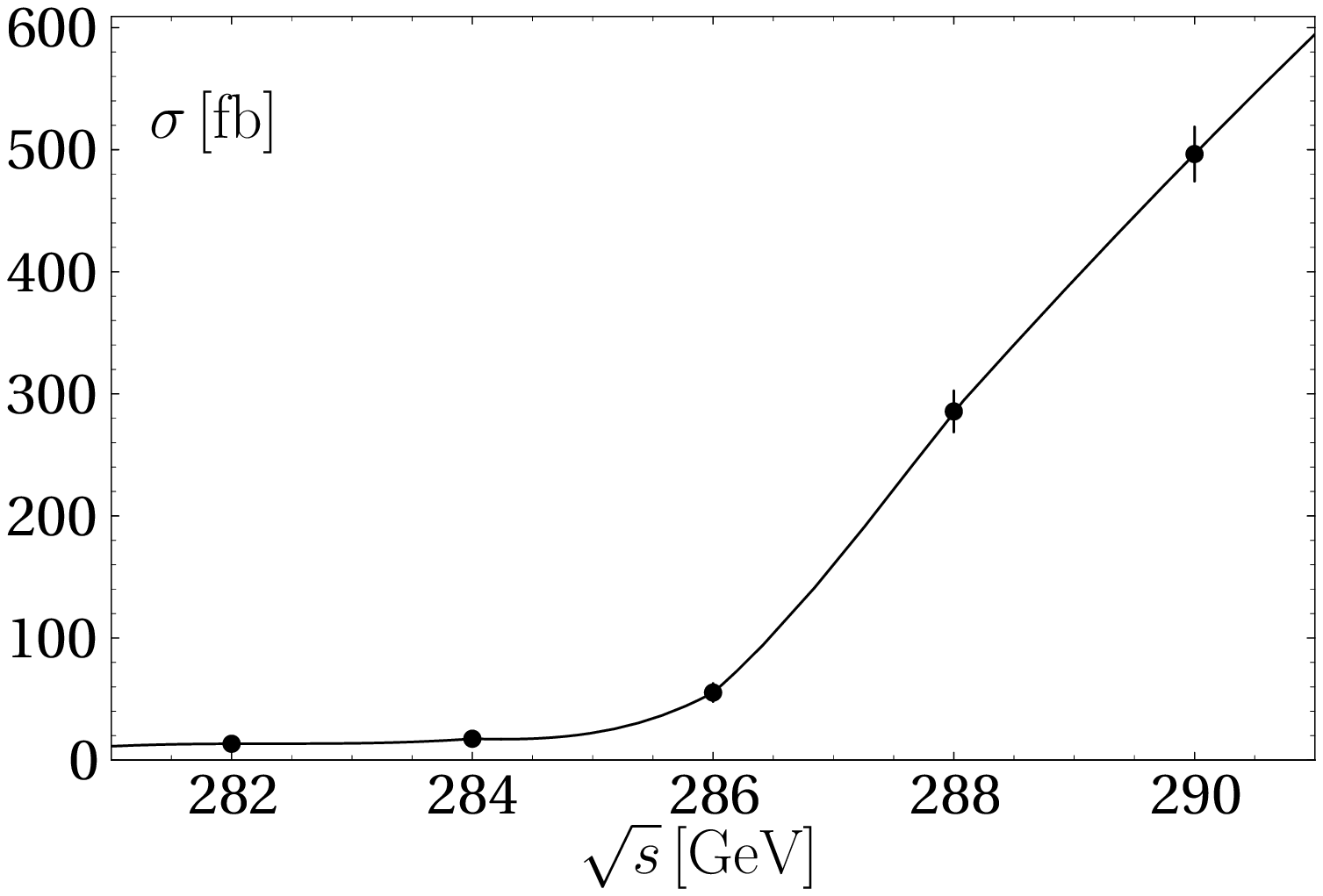}%
~~~~~\begin{minipage}[b]{50mm}
\caption{\it Cross sections at threshold for the reactions
   $e^+_Le^-_R\to\ser^+\ser^-$ (left) and
  $e^-_Re^-_R\to\ser^-\ser^-$ (right) in the SPS1a scenario,
  including background.
  Error bars correspond to a luminosity
  of 10 fb$^{-1}$ (left) and  1 bf$^{-1}$ (right) per point.\label{scans}. Note different vertical   scales.}
\end{minipage}
\end{figure}

At the ILC two methods can be used to measure sparticle masses:  threshold scans or in continuum.
The shape of
the production cross section near threshold is sensitive
to the masses and quantum numbers. For first 2 generations, where $R$-$L$ mixing can be neglected for example,
$\tilde{\mu}^+_L\tilde{\mu}^-_L\, , \,
\tilde{\mu}^+_R\tilde{\mu}^-_R\, ,
\sel^+ \sel^- $ and $ \ser^+ \ser^-$ pairs are excited in
P-wave characterized by a slow rise of the cross section
$\sigma \sim \beta^3$ with slepton
velocity $\beta$. On the other hand, in
$e_L^+ e_L^- \,/\, e_R^+ e_R^- \to \ser^+ \sel^- \,/\, \sel^+
\ser^-$ and $e_L^- e_L^- \,/\, e_R^- e_R^- \to \sel^- \sel^- \,/\,
\ser^- \ser^-$
sleptons are excited in  S-wave  giving steep
rise of the cross sections $\sigma\sim \beta$. Simulations for the SPS1a point\cite{SPS},
Fig.~\ref{scans}\cite{Freitas:2002gh}, show that   the ${\ser}$ mass can be determined
to 2 per mil; the resolution deteriorates by a
factor of $\sim2$ for $\tilde{\mu}^+_R\tilde{\mu}^-_R$ production.  For
$e^-_Re^-_R\to\ser\ser$ the fast rise of the cross section allows to gain a factor $\sim 4$
in precision already at a tenth of the luminosity if the $e^+e^-$ case.

Above the threshold, slepton masses can be obtained from the endpoint
energies of leptons coming from slepton decays. In the case of
two-body decays, $
   \tilde{\ell}^-  \to  \ell^-\nt_i \ $ and
  $   \tilde\nu_\ell  \to  \ell^-\tilde\chi^+_i $
the lepton energy
spectrum is flat with endpoints (the minimum $E_-$ and maximum $E_+$
energies) given by
\begin{eqnarray}
  E_{\pm}  & = &
       {\textstyle \frac{1}{4}} \sqrt{s} \, (1 \pm \beta)
        ( 1 - {m_{\cx}^2}/{m_{\tilde\ell}^2} )
  \label{eminmax}
\end{eqnarray}
Unlike at the LHC, the knowledge of the collision energy allows not only  an accurate determination of
the mass of the primary slepton but also the secondary neutralino/chargino.
One finds that $m_{\ser}$, $m_{\smur}$
and $m_{\nt_1}$ can be measured to $0.1$ to $0.18$ GeV, \ie\ 2 per mil in  selectron and
smuon production processes\cite{Martyn:2004ew}.
The  $\smul$ is more difficult to detect because of large
background from $WW$ pairs and SUSY cascades. However,   high
luminosity allows one to  select the rare decay modes
$\smul \to \mu \nt_2$ and  $\nt_2  \to  \ell^+ \ell^-\,\nt_1$
leading to a unique, background free
signature $\mu^+\mu^-\,4 \ell^\pm \not \!\!E$.
The achievable mass resolution for $m_{\smul}$ and $m_{\nt_2}$ is of the order
4 per mil\cite{Martyn:2003av}.

The chargino
masses can be measured very precisely at  threshold: simulations
for the reaction
$e^+_R e^-_L \to \cx^+_1 \cx^-_1 \to \ell^\pm \nu_\ell\cx^0_1\,
q\bar q' \cx^0_1$ show that the mass resolution is
excellent of ${\cal O}$(50~MeV), degrading to the per~mil level for the
higher $\cx^\pm_2$ state. Above threshold, from
the di-jet energy distribution in $\cx^\pm_2\to q\bar q'\nt_1$ one
expects a mass resolution of $\delta m_{\cx_1^\pm}=0.2$~GeV,
while the di-jet mass distributions constrains the
$\cx^\pm_1 - \nt_1$ mass splitting to about 100~MeV.
Similarly,
the di-lepton energy and mass distributions in the reaction
$e^+e^-\to\nt_2 \nt_2\to  4\ell^\pm \not \!\! E$ can be used to
determine $\nt_1$ and $\nt_2$ masses to about
2~per~mil\cite{Martyn:2003av}.
Higher resolution of order 100~MeV for $m_{\nt_2} $ can be obtained
from a threshold scan of $e^+e^-\to\nt_2\nt_2$; heavier states  $\nt_3$ and
$\nt_4$,
if accessible, can still be resolved with a resolution of a
few hundred MeV.

\subsection{Couplings and mixings}
The $L$-$R$ mixing for the
third generation can be non-negligible due to the large
Yukawa coupling making the $\tilde{\tau}$, $\tilde{t}$ and $\tilde{b}$ systems
very interesting to study to determine their mixing and chiral quantum
numbers. Likewise, we would like  to determine the
gaugino and higgsino composition of charginos and neutralinos. Equally important is  to verify the SUSY mass relations and exact equality
(at  tree level) of gauge couplings and their supersymmetric Yukawas.
For all these measurements the ability of having {\it both} beams, positrons
and electrons, polarised turns to be crucial\cite{Moortgat-Pick:2005cw}, 
since for many measurements  even  100\% electron polarisation is
insufficient.

\begin{figure}[h!]
\centering
\includegraphics[height=.17\textheight,width=.24\textwidth]{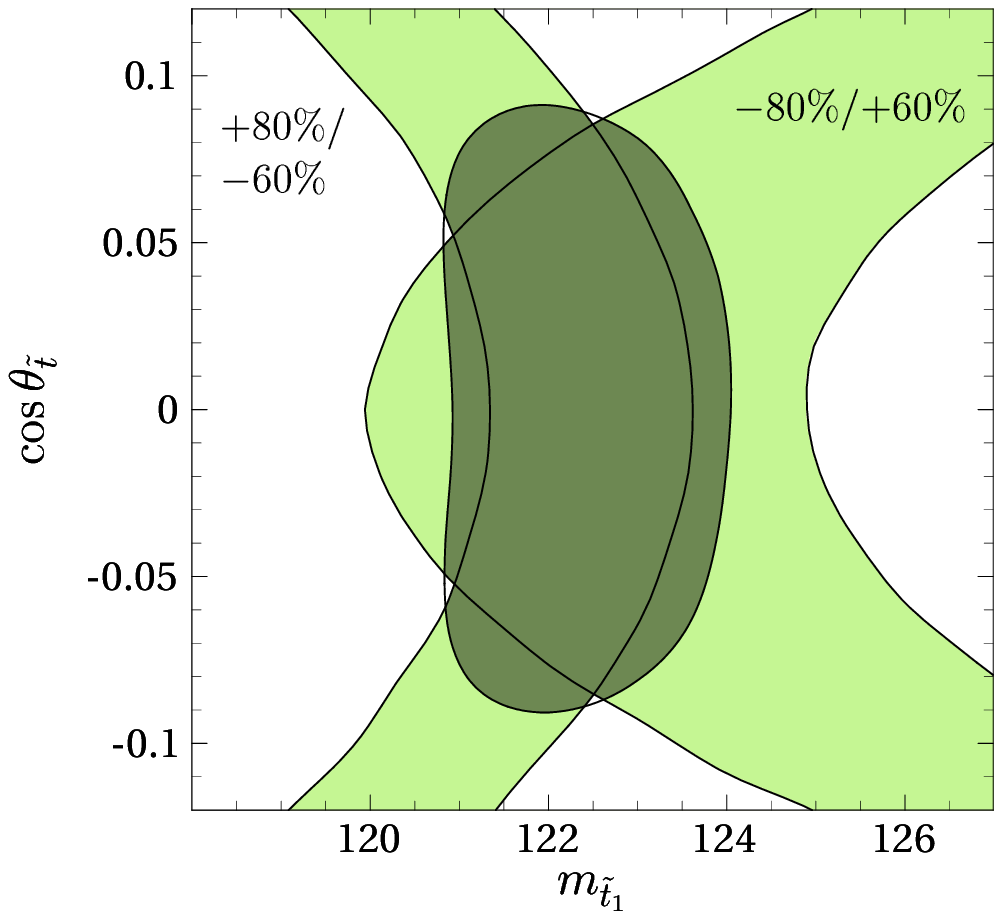}
\includegraphics[height=.17\textheight,width=.24\textwidth]{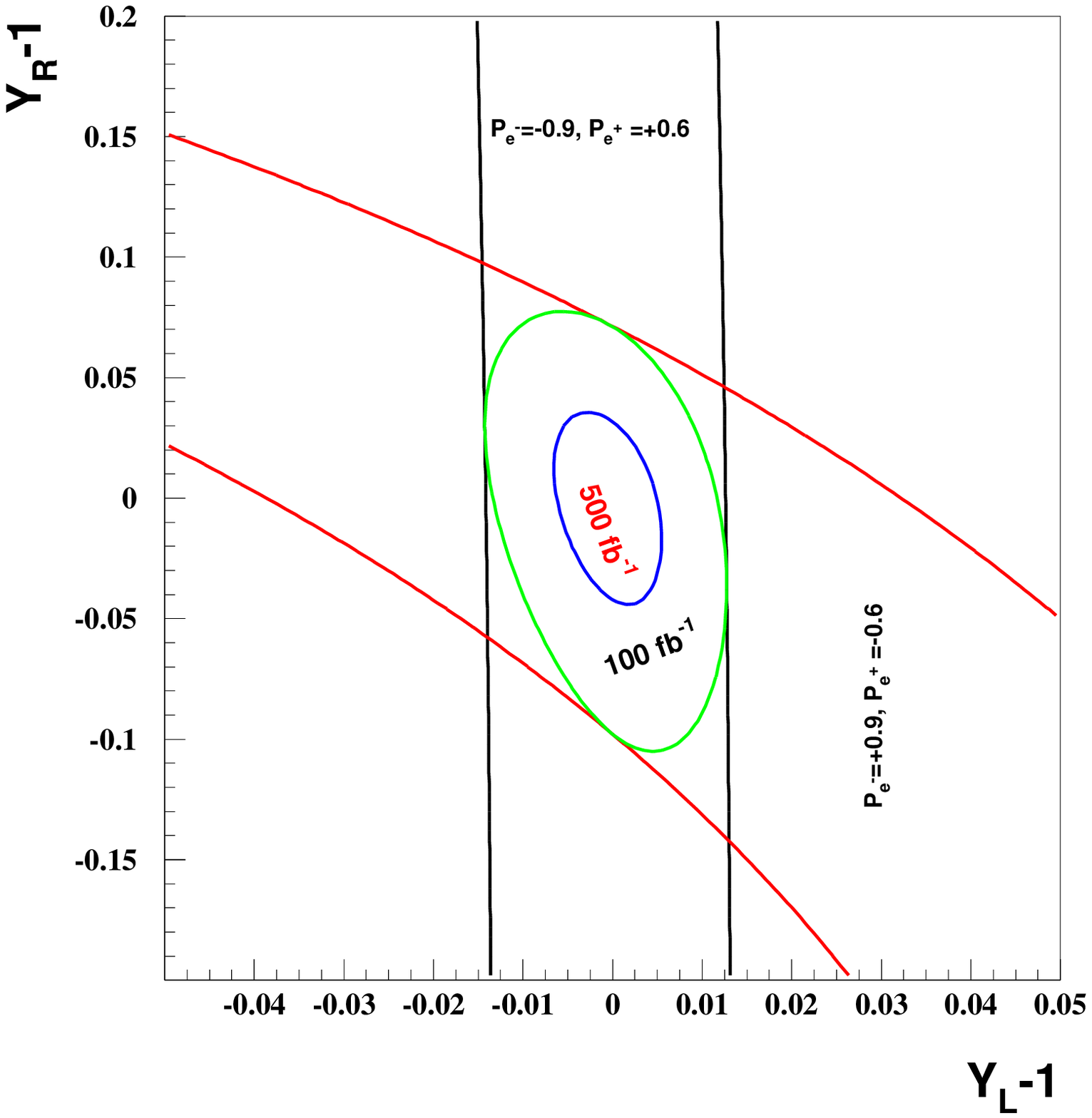}
\includegraphics[height=.17\textheight,width=.24\textwidth]{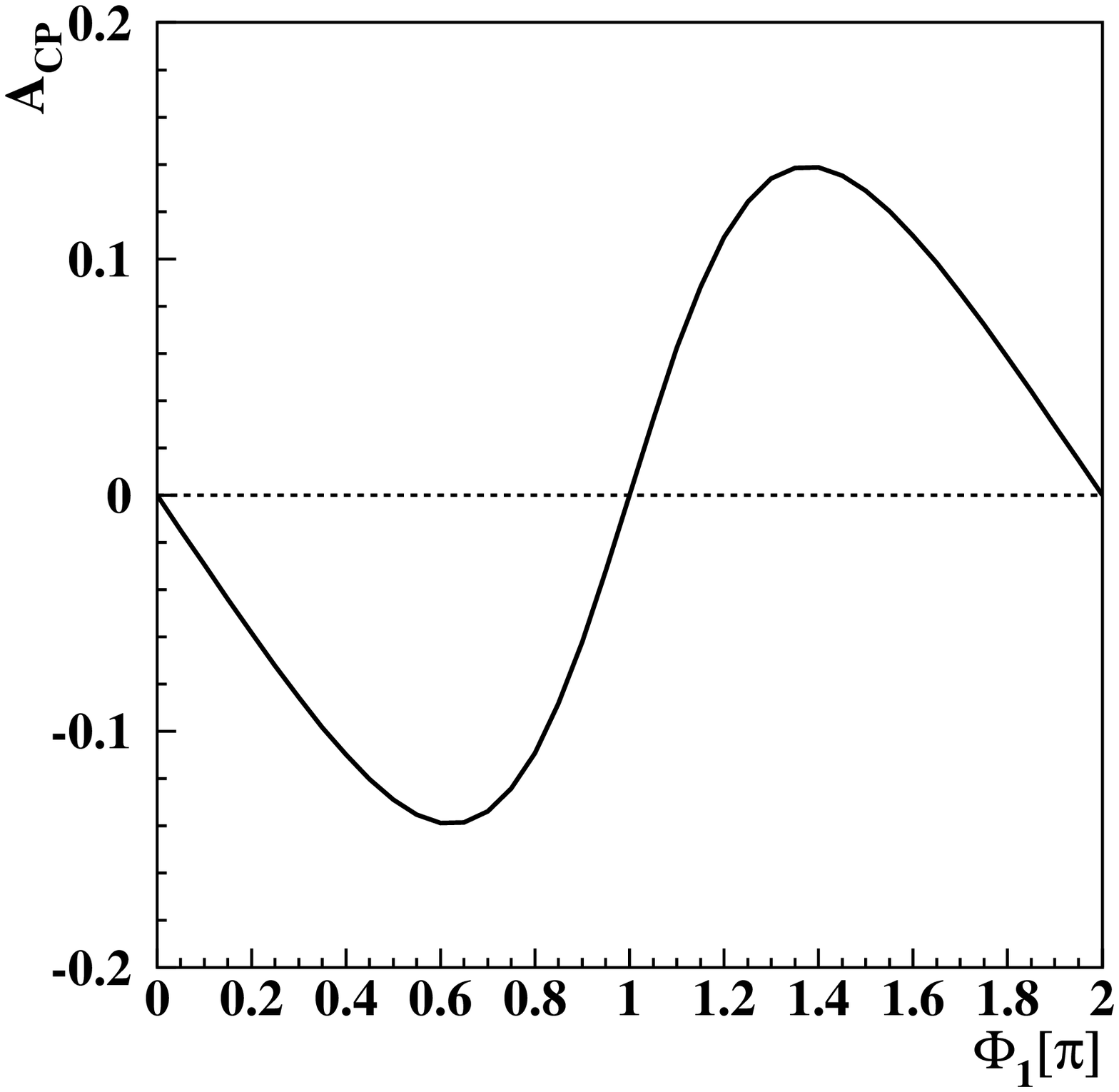}\\
\caption{\it Power of polarization -- bounds on: (left)
light stop mass $m_{\tilde{t}_1}$ and stop mixing angle
$\theta_{\tilde{t}}$ from $\sigma(e^+e^- \to
\tilde{t}_1 \tilde{t}_1^*)$;
(center) on $Y_L=\hat{g}_2/g_2$ and
       $Y_R=\hat{g}_1/g_1$ from
neutralino pair--production with
polarized beams. (right) $\Phi_1$ dependence of the CP--odd
  asymmetry $A_{\rm CP}$.
\label{fig:couplings} }
\end{figure}

The couplings and
mixing angles can be extracted from production cross sections
measured with polarized beams. For example, experimental analyses of stop
quarks with
small stop-neutralino mass difference,  motivated by the stop-neutralino
co-annihilation DM scenario, are very demanding. Nevertheless, the stop parameters can be
determined precise enough, Fig.~\ref{fig:couplings} (left), and
precisions for the dark matter predictions comparable  to that from direct
WMAP measurements in
the region  down to mass differences $\sim {\cal{O}}$(5 GeV)
can be achieved\cite{Carena:2005gc}.

The Yukawa couplings of scalar fermions can precisely be determined
by measuring the production cross-sections with polarized beams.
For example, in the electroweak sector, the relation between
the hypercharge
U(1)$_{\rm Y}$ coupling
$g_1$ and the SU(2)$_{\rm L}$ coupling $g_2$ and the corresponding
Yukawa couplings $\hat{g}_1$ and
$\hat{g}_2$ can
accurately be checked in neutralino pair--production.  Combining the
measurements of $\sigma_R$ and $\sigma_L$ for the process
$e^+e^-\rightarrow \tilde{\chi}^0_1\, \tilde{\chi}^0_2$, the
Yukawa couplings can be
determined to quite a high precision, as demonstrated in
Fig.~\ref{fig:couplings} (center)\cite{Choi:2001ww}.

Polarisation is a very powerful tool not only for preparing the desirable
initial state, but also as a diagnosis tool of final states. For example,
neutralinos $\tilde{\chi}^0_2$
produced in $\tilde{e}^\pm_L$ decays are 100\% polarized \cite{Aguilar1}.
Furthermore, in $e^+e^-\to\tilde{e}^+_L\tilde{e}^-_L
\to e^+\tilde{\chi}^0_1 e^-\tilde{\chi}^0_2$
followed by the
three--body decay $\tilde{\chi}^0_2\to \tilde{\chi}^0_1\mu^+\mu^-$
it is possible to reconstruct the rest frame of the
neutralino $\tilde{\chi}^0_2$ as shown in Ref.\cite{Aguilar23}.
Such a perfect
neutralino polarization combined with the study of angular correlations
in the neutralino rest frame  can provide us with
ways for  probing the Majorana nature of the
neutralinos and CP violation in the neutralino
system. With the neutralino spin vector $\hat n$
and two final lepton momentum directions $\hat q^+$ and $\hat
q^-$ the CP--odd asymmetry
can be
constructed by comparing number of
 events with $O_{\rm CP}=\hat{n}\cdot(\hat{q}_+\times\hat{q}_-)$ positive and
 negative, normalized to the sum. Fig.~\ref{fig:couplings} (right) shows the
 dependence of the CP-odd asymmetry on the phase $\Phi_1$
of the bino mass parameter
 $M_1$\cite{Choi:2005gt}.

\subsection{Looking beyond the ILC kinematic reach}

The precision measurements offered by the ILC allow us to infer indirect
information on heavy states not directly accessible. As an illustration we
consider two examples.

\begin{figure}[htb]
\includegraphics*[width=45mm,height=35mm]{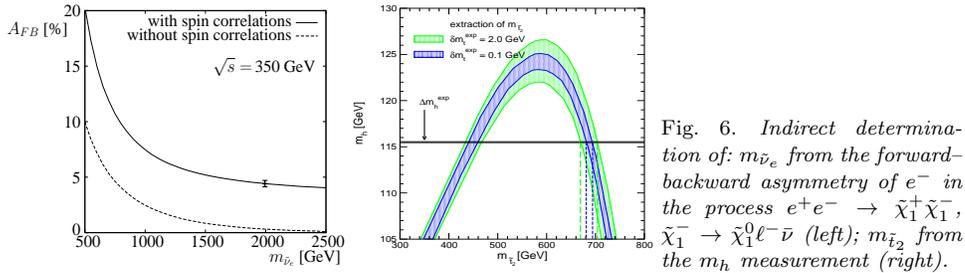}
\includegraphics*[width=40mm,height=35mm]{mhMSt202b.cl.eps}
\begin{minipage}[b]{4cm}
\caption{\it Indirect determination of:
$m_{\tilde{\nu}_e}$ from the
 forward--backward asymmetry of $e^-$ in
the process $e^+e^- \to \tilde{\chi}^+_1 \tilde{\chi}^-_1$,
$\tilde{\chi}^-_1 \to \tilde{\chi}^0_1 \ell^- \bar{\nu}$ (left);
$m_{\tilde{t}_2}$
from the $m_h$ measurement
(right).
\label{fig:beyond} }
\end{minipage}
\end{figure}

The first example concerns an interesting scenarios in which scalar sparticle sector is heavy
while the gaugino masses are kept relatively small,  like in the cosmology-motivated
focus-point scenario \cite{focuspoint}.
Precision analyses of cross sections for light
chargino production and forward--backward asymmetries of decay leptons at the
first stage of the ILC, Fig.~\ref{fig:beyond} (left), together with mass
information on ${\tilde{\chi}^0_2}$ and squarks from the LHC, show that the
underlying fundamental gaugino/higgsino MSSM parameters and constrains on the
heavy, kinematically inaccessible sparticles with masses $\cal O$(2 TeV), can
be obtained nevertheless\cite{Desch:2006xp}.

If the second top squark $\tilde{t}_2$ is too heavy for the ILC, and due to huge
background invisible at the LHC, the precise measurement of the Higgs boson
mass $m_h$ at ILC together with measurements from the LHC can be used to obtain
indirect limits on $m_{\tilde{t}_2}$\cite{Heinemeyer:2003ud}, Fig.~\ref{fig:beyond}.
Both  examples again demonstrate the power of the LHC/ILC interplay,
since neither of these colliders alone can provide sufficient data needed
to determine the SUSY parameters in such difficult scenarios.

\subsection{$e^-e^-$, $e\gamma$ and $\gamma\gamma$ options}

Compton back-scattering of the laser light on  electron beam(s) opens a
possibility of converting the $e^-e^-$ collider to an $e\gamma$ and
$\gamma\gamma$ collider with energies and luminosities comparable to those of
$e^+e^-$ collider\cite{Telnov:2006cj}. If realized, these options may open
new  discovery channels. Again I will take two specific
examples to illustrate the
point.

\begin{figure}[htb]
\includegraphics*[width=40mm,height=35mm]{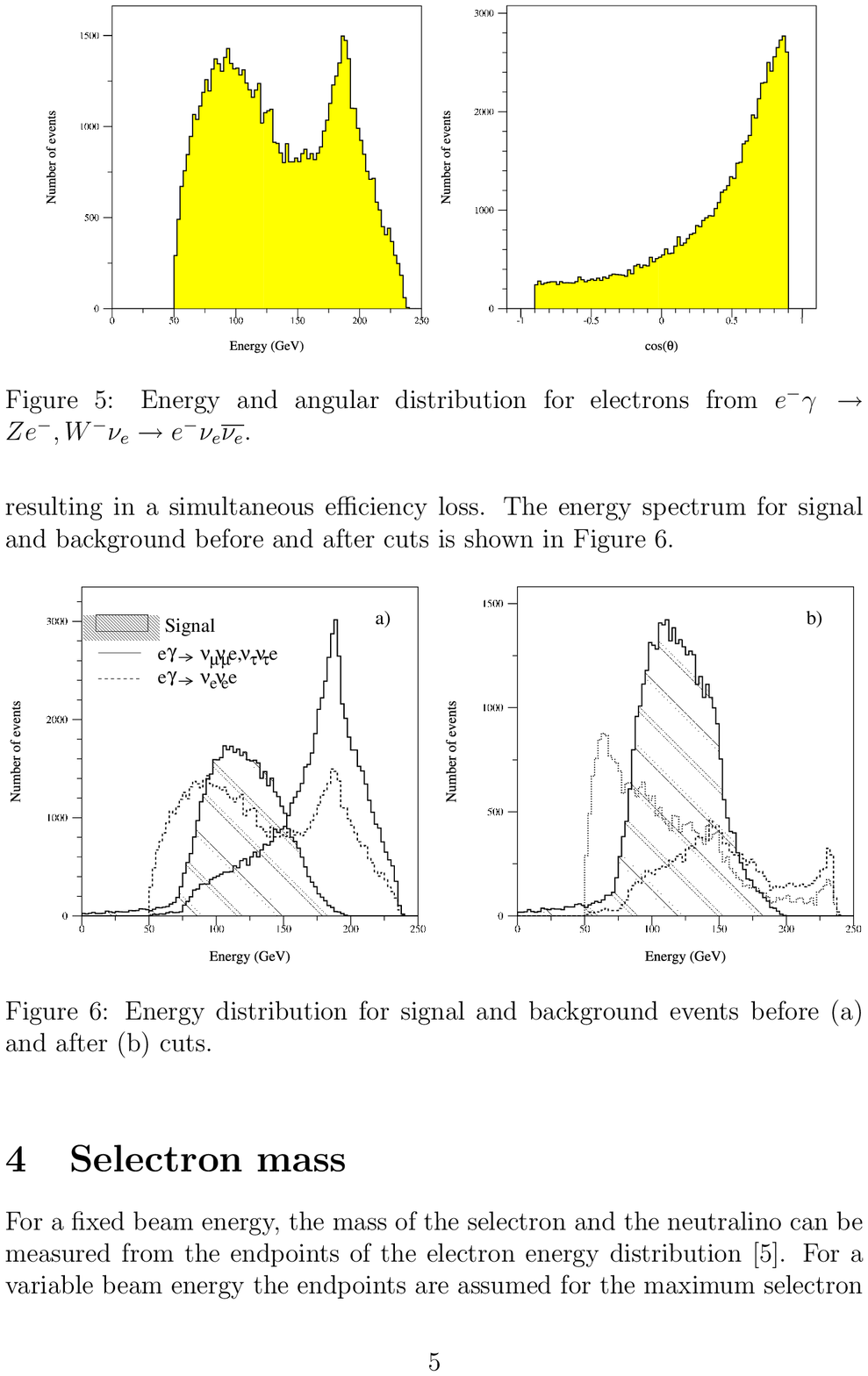}
\includegraphics*[width=40mm,height=35mm]{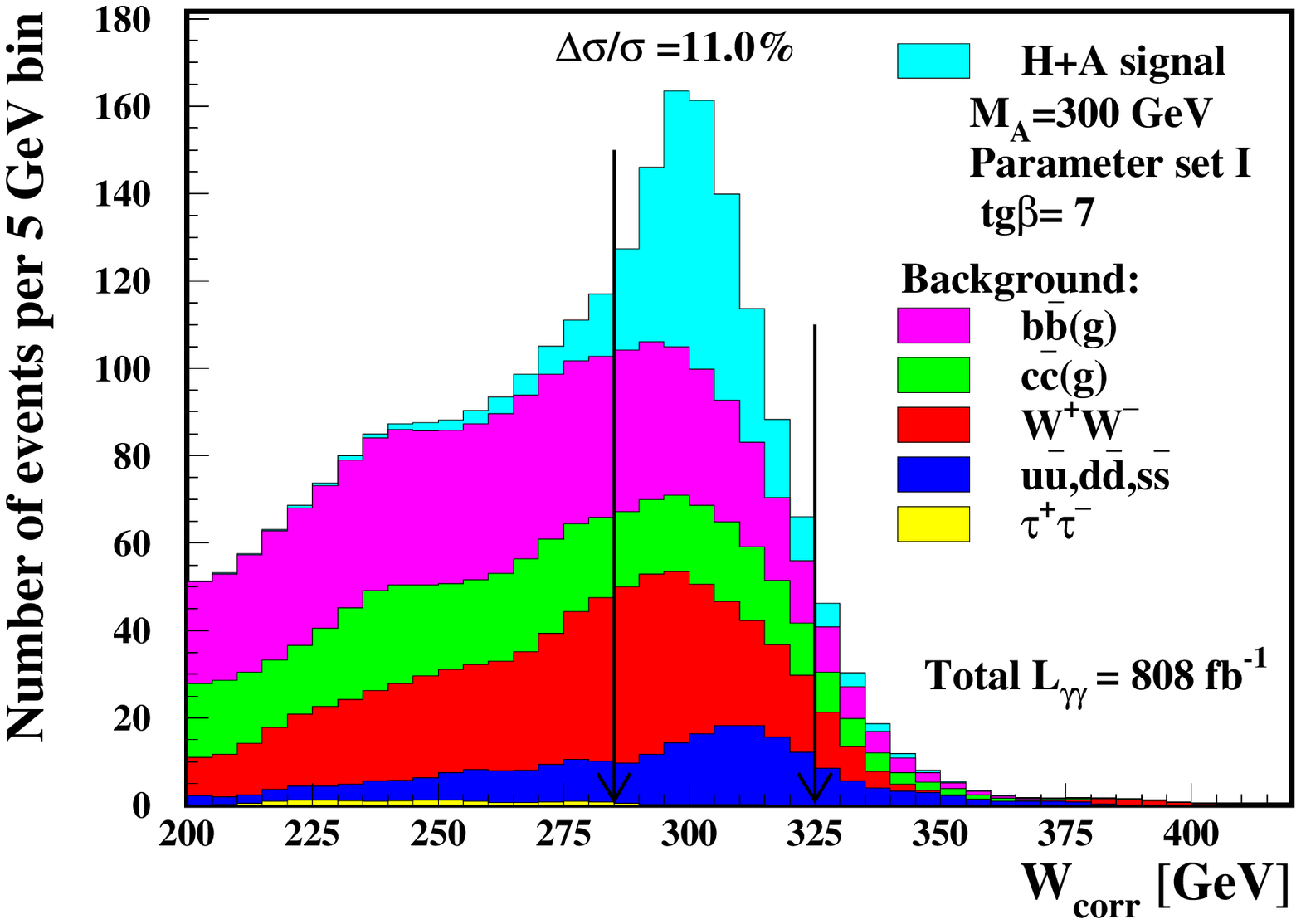}~~\begin{minipage}[b]{4cm}
\caption{\it Electron energy distribution signal $e\gamma\to \nu_e\nu_e e$
  (shaded) and background after cuts (left).
Invariant mass distribution for $\gamma\gamma\to H,A\to b\bar
b$ and and background (right). \phantom{and and background (right)} \label{fig:options}}
\end{minipage}
\end{figure}

If the mass difference between the
lightest neutralino and the selectron is a few hundred GeV, it may
happen that chargino pair production at the ILC is possible, while
selectron pair production is kinematically forbidden. However,
$m_{\tilde{\chi}^0_1}+m_{\tilde{e}}$ can still be below 90\% of the
centre-of-mass energy, so that the process
$e\gamma\to \tilde{\chi}^0_1\, \tilde{e}^-$
is possible at an $e\gamma$  collider. If the photon energy  were known,
the selectron and neutralino masses could be determined from the endpoints of
the decay electron
distribution, like in $e^+e^-$ collisions. Although the variable photon energy
smears the endpoints, simulations have shown (Fig.~\ref{fig:options}) that
with the  $m_{\tilde{\chi}^0_1}$ determined in $e+e^-$ running,
the selectron mass can be  reconstructed from the position of the 
lower edge\cite{AlvarezIllan:2005xs}.  

$\gamma\gamma$ collider offers a unique possibility of producing  as
$s$-channel resonances neutral
Higgs bosons $H,A$ that are {\it both}
too heavy to be produced in associated $HA$ or
$ZH$ processes at $e^+e^-$ collider {\it and}
lay in the so called ``LHC-wedge'' of
intermediate  values of $\tan\beta$, to
which the LHC is blind. Results of a simulation for the combined
$\gamma\gamma\to H,A\to b\bar
b$ analyses are shown in Fig.~\ref{fig:options}\cite{Niezurawski:2006ia} 
(the $H$ and $A$ bosons are almost
mass-degenerate). Other decay modes ($WW$, $ZZ$,
$t\bar t$) can provide a
means to determine the Higgs-boson CP properties \cite{Niezurawski:2004ga},
and the $\tau$-fusion process, $\gamma\gamma\to \tau\tau H,A$, can serve to
measure $\tan\beta$\cite{Choi:2004ne},
the parameter that is notoriously difficult to determine
experimentally.

\subsection{Beyond the ILC}
It is expected that
higher energy colliders will be needed to help unravel the multi-TeV physics
left unveiled either by the LHC or by the ILC. Further progress in particle
physics may require clean experiments at a linear $e^+e^-$ collider
at multi-TeV energies, like
CLIC\cite{Accomando:2004sz}, which would be an ideal machine to complement the
the LHC and ILC physics program.  Simulations for CLIC concentrated
on such scenarios with
sparticles beyond the LHC and ILC reach.

\begin{figure}[htb]
\includegraphics*[width=40mm,height=35mm]{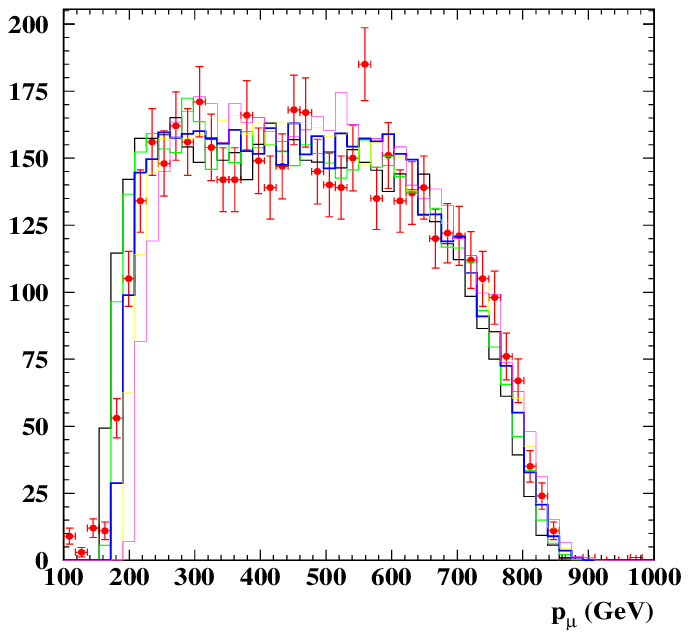}
\includegraphics*[width=40mm,height=35mm]{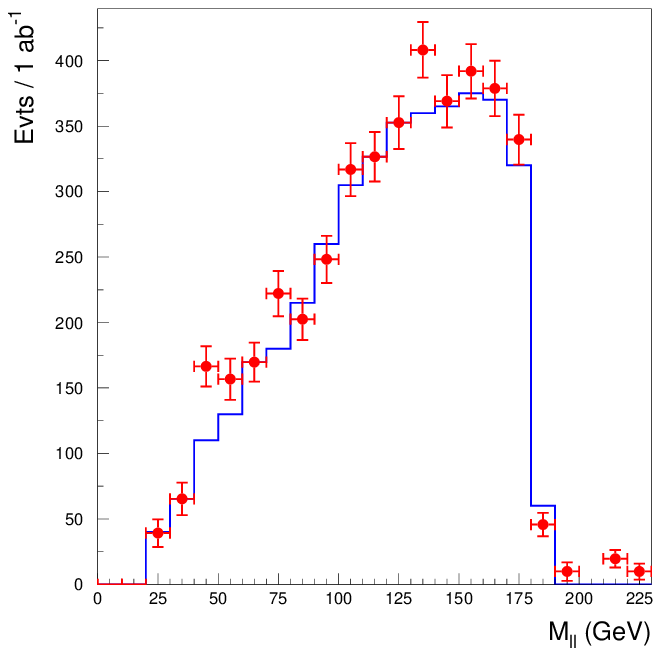}
\includegraphics*[width=40mm,height=35mm]{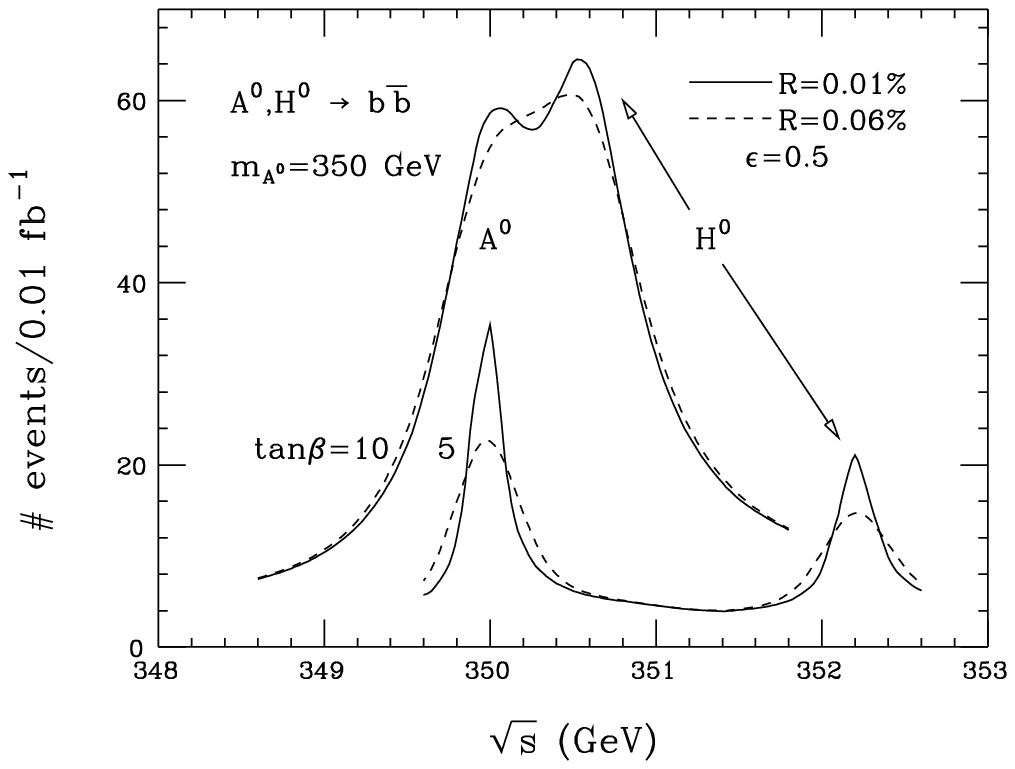}~~~
\caption{\it Muon energy spectrum from
  $\tilde{\mu}_L\to\mu\tilde{\chi}^0_1$ (left), and
di-muon invariant mass spectrum from $\tilde{\chi}^0_2\to
 \mu^+\mu^-\tilde{\chi}^0_1$ (middle) at CLIC.
Separation of $A$ and $H$ signals at a muon
collider (right).}
\label{fig:beyondILC}
\end{figure}

Fig.~\ref{fig:beyondILC} (left) shows simulations of the muon energy spectrum
from a 1150 GeV selectron decaying to a muon and a 660 GeV LSP neutralino.
The endpoints are clearly seen allowing the selectron and
neutralino mass determination. Likewise, in  Fig.~\ref{fig:beyondILC}(middle)
the di-muon invariant mass distribution from $\tilde{\chi}^0_2\to
 \mu^+\mu^-\tilde{\chi}^0_1$ exhibits a pronounced edge which, together
with results from selectron decay
make a measurement of $m_{\tilde{\chi}^0_2}$ possible.

In more distant future a muon collider with extremely good beam energy
resolution will provide a tool
to explore Higgs (and Higgs-like objects)  by direct $s$-channel fusion because of
enhanced couplings of muons to Higgs bosons, much
like the LEP explored the $Z$. Right panel of
Fig.~\ref{fig:beyondILC} demonstrates how well
two almost mass-degenerate Higgs bosons $H$ and $A$ can be resolved
\cite{Alsharoa:2002wu}.

\section{Reconstructing the underlying SUSY model}

The expected high experimental accuracies at the LHC/ILC
could not be fully exploited if not  matched from the theoretical
side. This calls for  a well defined
theoretical framework for the calculational schemes in
perturbation theory as well as for the input parameters. Motivated by the
experience   in analyzing data
at the former $e^+e^-$ colliders LEP and
SLC, and building on vast experience in SUSY calculations and data simulations and
analyses, the Supersymmetry Parameter Analysis (SPA)
Convention and Project\cite{SPA} has been proposed. It recommends
a convention for high-precision theoretical calculations, and provides
a program repository of numerical codes,
a list of tasks needed further improvements and
a SUSY reference point SPS1a$'$ as a test-bed.

The SPA Convention and Project is a joint inter-regional   effort that could
serve as a forum to discuss future improvements on both experimental and
theoretical sides to exploit fully the physics potential of LHC, and ILC.
The current status of the project
is documented on  the web-page {\tt  http://spa.desy.de/spa/
}

\subsection{SPA Convention}
The SPA Convention consists of the following propositions:\\[1mm]
\phantom{a}$\bullet$  \begin{minipage}{.95\textwidth} The masses of the SUSY particles and Higgs bosons are
defined as pole      masses.\end{minipage}\\[1mm]
\phantom{a}$\bullet$ \begin{minipage}[t]{.95\textwidth} All SUSY Lagrangian parameters, mass parameters and
couplings,
     including
     $\tan\beta$, are given in the $\drbar$ scheme at the scale
     $\tilde M =$ 1 TeV.\end{minipage}\\[1mm]
\phantom{a}$\bullet$ \begin{minipage}[t]{.95\textwidth} Gaugino/higgsino and scalar mass matrices, rotation
matrices and the
     corresponding angles are defined in the $\drbar$ scheme at $\tilde M$,
     except for the Higgs system in which the mixing matrix is defined in the
     on-shell scheme, the scale parameter chosen as the light Higgs mass.\end{minipage}\\[1mm]
\phantom{a}$\bullet$  \begin{minipage}[t]{.95\textwidth} The Standard Model input parameters of the gauge sector
are chosen as
     $G_F$, $\alpha$, $M_Z$ and $\alpha_s^{\msbar}(M_Z)$. All lepton masses are
     defined on-shell. The $t$ quark mass is defined on-shell; the $b,\, c$
     quark masses are introduced in $\msbar$ at the scale of the
     masses themselves while taken at a renormalization scale of 2 GeV for
     the light $u,\, d,\, s$ quarks.\end{minipage}\\[1mm]
\phantom{a}$\bullet$ \begin{minipage}[t]{.95\textwidth} Decay widths, branching ratios and production cross
sections are
     calculated for the set of parameters specified above.\end{minipage}

\subsection{Program repository}

The repository contains links to codes grouped in several categories: scheme
translation tools; spectrum calculators from the Lagrangian parameters;
calculators of various observables: decay tables, cross sections, low-energy
observables, cold dark matter relics, cross sections for CDM particle
searches; event generators; analysis programs to extract the Lagrangian
parameters from experimental data; RGE codes; as well as some auxiliary
programs and libraries.

The responsibility for developing codes and maintaining them up to the current
theoretical state-of-the-art precision rests with the authors. The SLHA
\cite{Skands:2003cj} convention is recommended for communication between the
codes.

\subsection{The test-bed: Ref.\ Point SPS1a$'$ }

To perform first checks of its internal consistency and to explore
the potential of such coherent data analyses a MSSM Reference Point SPS1a$'$
has been proposed as a testing ground.
The roots defining SPS1a$'$ are the mSUGRA
parameters
$ M_{1/2} = 250$ GeV, $M_0 =  70$ GeV, $A_0=-300$ GeV  at the GUT
scale, and $\tan\beta (\tilde M)  = 10$, $\mu>0$.
The point is
close to the original Snowmass point SPS1a\cite{SPS} and
to point $B'$ of
\cite{Battaglia:2003ab}.
 Recently global analysis programs have become available
\cite{Lafaye:2004cn} in which the whole set of data, masses, cross sections,
branching ratios \etc, is exploited coherently to extract the Lagrangian
parameters in the optimal way after including the available radiative
corrections.

The parameter set SPS1a$'$ chosen  for a first study
   provides a benchmark  for
   developing and testing the tools needed for a successful analysis
   of future SUSY data.  However, neither this specific point nor
   the MSSM itself may be the correct model for low-scale SUSY.
Other scenarios might be realized
   in the SUSY sector and the SPA convention is
   general enough to cover them.

Although current SPA studies are very encouraging,
much additional work both on
the theoretical as well as on the experimental side will be needed to achieve
the SPA goals.

\section{Summary}

Much progress has been achieved in  preparing the physics programme for new machines.
At the beginning the LHC
has been considered merely as a discovery machine. However, over the years
many techniques
have been developed for extracting masses and couplings, and in some cases the
Lagrangian parameters.
Many  experimental analyses are still
based on lowest--order expressions.
On the theory side many higher-order calculations have been completed and
implemented in numerical codes. New theoretical ideas
deserve experimental analyses. However, the task of exploring all masses and
couplings of SUSY particles is probably impossible by the LHC
alone. The ILC will extend the discovery reach, in particular in the
electroweak sector, and greatly improve on precision SUSY measurements.
We still need new ideas and
techniques to explore fully the opportunities offered to us by the LHC and
ILC.
The SPA Convention and Project
should prove very useful in streamlining discussions and
comparisons of different calculations and experimental analyses.

\vspace*{12pt}
\noindent
{\bf Acknowledgment}
\vspace*{6pt}

\noindent
I would like to thank the organizers and Dr. Khalil Shaban in particular for
hospitality and creative atmosphere at the conference.
Work supported by
the Polish Ministry of Science and Higher Education Grant No~1~P03B~108~30.


\end{document}